\documentclass[english]{cccconf}

\usepackage{algorithm}
\usepackage{algpseudocode}
\usepackage{amsmath}
\usepackage{graphics}
\usepackage{epsfig}

\usepackage{bm}

\usepackage[comma,numbers,square,sort&compress]{natbib}
\usepackage{epstopdf}

\usepackage{graphicx}
\usepackage{dcolumn}
\usepackage{bm}

\usepackage{amsmath} 
\usepackage{esint}
\usepackage{amsfonts}
\usepackage{mathrsfs} 
\usepackage{multirow}

\usepackage{color}

\usepackage[colorlinks,
linkcolor=blue,
anchorcolor=blue,
citecolor=blue]{hyperref}

\begin{document}

\title{An unsupervised feature learning for quantum-classical convolutional network with applications to fault detection}

\author{Tong Dou\aref{SCUT},
        Zhenwei  Zhou\aref{MIIT},
         Kaiwei Wang\aref{MIIT},
         Shilu Yan\aref{SCUT},
         Wei Cui\aref{SCUT}}

\affiliation[SCUT]{School of Automation Seience and Engineering, South China University of Technology, Guangzhou 510641, China
        }
\affiliation[MIIT]{The Fifth Institute of Electronics, Ministry of Industry and Information Technology, Guangzhou 511370, China
 \email{zhouzw\_ceprei@163.com,~~wangkaiwei82@163.com,~~aucuiwei@scut.edu.cn}
}





\maketitle

\begin{abstract}
Combining the advantages of quantum computing and neural networks, quantum neural networks (QNNs) have gained considerable attention recently. However, because of the lack of quantum resource, it is costly to train QNNs. 
In this work, we presented a simple unsupervised method for quantum-classical convolutional networks to learn a hierarchy of quantum feature extractors. Each level of the resulting feature extractors consist of multiple quanvolution filters, followed by a pooling layer. The main contribution of the proposed approach is to use the $K$-means clustering to maximize the difference of quantum properties in quantum circuit ansatz. One experiment on the bearing fault detection task shows the effectiveness of the proposed method.
\end{abstract}


\footnotetext{This work was supported by the National Natural Science Foundation of China under Grant 61873317, and in part by Guangdong Basic and Applied Basic Research Foundation under Grant 2020A1515011375.  \textit{Corresponding authors: Zhenwei Zhou, Kaiwei Wang, and Wei Cui}}

\section{Introduction}
Quantum computers are devices that harness the laws of quantum physics, such as superposition and entanglement, to perform computation. Benefiting from super parallel computing power in principle, quantum computers are expected to slove certain problems that classical computers either cannot solve, or not solve in any reasonable amount of time. The growth of computing power and the rapidly increasing volume of data make a great progress of machine learning (ML) teachniques that build complex models for finding patterns in data. As  the data processing ability of classical computers is approaching the limit, quantum computing is believed to promote the development of machine learning because of its powerful parallel computing power. The intersection between machine learning and quantum computing, called quantum machine learning (QML)\cite{biamonte2017quantum}, has attracted more and more attention in recent years. The goal of quantum machine learning is to take advantages of quantum properties to achieve better performance than the classical machine learning teachniques in terms of computational complexity or pattern recognition. This has led to a number of quantum machine learning algorithms\cite{lloyd2014quantum,PhysRevLett113130503,PhysRevX8021050}, such as qSVM, qPCA, quantum Boltzmann machine. Some of these algorithms are shown to be more effecient than their classical counterparts. However, it is hard to implement them in noisy intermediate scale quantum (NISQ)\cite{Preskill2018quantumcomputingin} devices which may include a few tens to hundreds of qubits without error correction capability.Recently, several NISQ algorithms which are based on parameterized quantum circuits (PQCs), such as vatiational quantum eigensolvers (VQE)\cite{peruzzo2014variational,PhysRevA.92.042303,mcclean2016the} for ground states, quantum approximate optimization algorithm (QAOA)\cite{farhi2014a} for combinatorial problems and quantum kernel methods\cite{havlivcek2019supervised,schuld2019quantum} for classification, have been developed.

In a way, PQCs offer a promising path for NISQ era. Based on PQCs, quantum neural networks(QNNs) \cite{farhi2020classification,mari2020transfer} have been proposed. As an important kind of model of classical neural networks, convolutional neural networks (CNNs), which are designed for processing data that has a known grid-like topology, are good at computer vision tasks, such as image recognition, image segmentation and objection detection. Utilizing the thoughts of CNNs, quantum convolutional neural networks (QCNNs) models\cite{cong2019quantum,henderson2020quanvolutional} are proposed.
In \cite{henderson2020quanvolutional}, authors replaced the convolution operation with the unitary transformation of a random quantum circuit by introducing a new quanvolutional layer which consists of quanvolutional filters. Similar to the conventional convolutional layer, a quanvolutional layer can be considered as a feature extractor. And the pooling layers and fully connected layers remain classically. Quanvolutional layers can easily integrate into classical CNNs to form a quantum-classical hybrid model, which can be implemented in near-term NISQ devices. However, using random quantum circuits with parameters unchange, the properties of circuits, such as expressibility and entangling capability\cite{sim2019expressibility}, will become more and more similar as the number of quanvolutional filters grows. On the other hand, quantum resources are scarce and expensive nowadays. A major drawback of many feature learning based QML algorithms is their complexity and expense because they need to be run many times to estimate the gradients. Thus we need a method to initialize the structure and the parameters of quanvolutional filters.

Based on the hybrid model introduced in \cite{henderson2020quanvolutional}, we proposed an unsupervised feature learning method to adress the problem metioned above in this paper. We use K-means algorithm to cluster the quantum circuits in different structures and parameters. And the quantum circuits which are closest to the cluster centers are initialized as quanvolutional filters. Once the quanvolutional filters are determined, we can then extract the features for the entire training set just once, essentially constructing a new training set for the last layer, which means that it is possible to use this unsupervised method to train quantum-classical models without ever using quantum resources during the training process.

This paper is organized as follows. Section II is the preliminary, in which we first provide a brief background of the framework of PQCs and the K-means clustering method. Then, the proposed unsupervised feature learning method for quanvolutional layers is described in detail in Section III. In Section IV, to verify the effiiency,case study on bearing fault detection is presented through numerical simulation. Conclusion are given in Section V. 


\section{Preliminaries}
In this section, we will breifly introduce the concepts of parameterized quantum circuits and K-means clustering method. 

\subsection{The Framework of Parametrized Quantum Circuits}
Parametrized quantum circuits(PQCs) are a kind of quantum circuits that have trainable parameters subject to iterative optimizations. In general, a PQC can be described as 
\begin{eqnarray}
U(\bm{\theta})=\prod_{j=1}^{M}U_{j}(\theta_{j}),
\end{eqnarray}
where $\bm{\theta}=(\theta_{1}, \dots, \theta_{M})$ are tunable parameters, while $U_{j}=e^{-i\theta_{j}V_{j}}$ is a rotation gate of angle $\theta_{j}$ generated by a Hermitian operator $V_{j}$ such that $V_{j}^{2}=\bm{I}$. In this paper, $V_{j}$ is one of Pauli matrices.

\begin{figure}[!htb]
  \centering
  \includegraphics[width=\hsize]{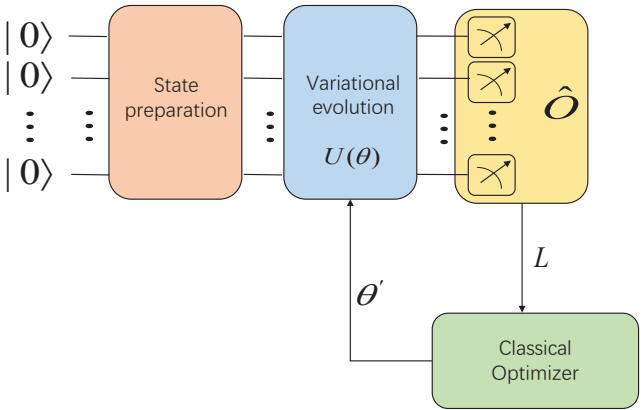}
  \caption{The illustration of three steps of PQCs based algorithms.}
  \label{fig1}
\end{figure}

Algorithms involving PQCs usually works in a hybrid quantum-classical scheme,  as shown in Fig.~\ref{fig1}. In general, a hybrid quantum-classical scheme combines quantum state preparation, variational evolution and measurement with classical optimization.

\begin{enumerate}
 \item First, prepare a intial state $|\varphi_0\rangle$ by encoding input into the quantum device.
 \item Second, we need to choose an appropriate ansatz  $U(\bm{\theta})$, that is, designing the circuit structure of a PQC, and apply $U(\bm{\theta})$ to $|\varphi_0\rangle$, where $\theta$ is parameters of the circuit.
 \item Then measure the circuit repeatedly on a specific observable $\hat{O}$ to estimate an expectation value $\langle\hat{O}(\bm{\theta})\rangle$. Based on the $\langle\hat{O}(\bm{\theta})\rangle$ which is fed into a classical optimizer, we compute a cost function $L( \langle\hat{O}(\bm{\theta})\rangle)$ to be minimized by updating $\bm{\theta}$. 
 \end{enumerate}
 
 These steps need to be repeated until when an approximate solution is reached.

\subsection{K-means}
The K-means method is a prototype-based objective function clustering method that selects the sum of the Eucilidean distances of all objects to the prototype as the objective function of the optimization. The problem is described mathematically as: Given a dataset $D = \{ \bm{x}_{1}, \bm{x}_2, \cdots, \bm{x}_{m} \}$ and a number $K$, find a partition of $K$ clusters to the dataset $D$ by optimizing the partitioning criterion:
\begin{eqnarray}
\min E = \sum_{k=1}^{K} \sum_{\bm{x} \in C_{k}} d^{2}(\bm{x}, \bm{\mu}_{k}),
\end{eqnarray}
where $\bm{\mu}_{k} = \frac{1}{|C_{k}|} \sum_{\bm{x}\in C_{k}} \bm{x}$ denotes the mean vector of $C_{k}$, and $d^{2}(\bm{x}, \bm{\mu}_{k}) = || \bm{x} - \mu_{k} ||_{2}^{2}$. To cluster all objects into $K$ classes, first select K initial particles randomly, assign each object to the particle with the smallest Euclidean distance to form $K$ clusters, and calculate the mean of each cluster as the new $K$ particles. Iterate continuously until the shutdown condition is met. In this way, one can easily classify all the objects into K classes. Concretely, the K-means algorithm can be described as follow.

\begin{algorithm}\label{ag1}
\caption{K-means Clustering Algorithm} 
\hspace*{0.02in} {\bf Input:} 
dataset $D = \{ \bm{x}_{1}, \bm{x}_{2}, \cdots, \bm{x}_{m} \}$;\\
\hspace*{0.4in} the number of clusters K.\\
\hspace*{0.02in} {\bf Output:} 
the clusters $C = \{  C_{1}, C_{2}, \cdots, C_{K} \}$.
\begin{algorithmic}[1]

\State randomly select $K$ samples from $D$ as initial mean vectors $\{ \bm{\mu_{1}}, \bm{\mu_{2}}, \cdots, \bm{\mu_{K}} \}$; 

\Repeat
\State  let $C_{i}=\emptyset$ $(1\leq i \leq K)$; 

\For{$j =1, 2, \cdots, m$}
\State compute the distance between $\bm{x}_{j}$ and each of mean vectors $\bm{\mu}_{i}(1 \leq i \leq K)$: $d_{ji} = || \bm{x}_{j} - \bm{\mu}_{i} ||_{2}$;
\State determine the cluster of $\bm{x}_{j}$ according to the distance of mean vectors: $\lambda_{j} = \arg \min_{i\in\{ 1,2,\cdots, K \}}d_{ji}$;
\State update the cluster $C_{\lambda_{j}}$: $C_{\lambda_{j}} = C_{\lambda_{j}} \cup \{ \bm{x}_{j} \}$;
\EndFor

\For{$i=1,2,\cdots,K$}
\State compute the new mean vectors: $\bm{\mu}_{i}^{\prime} = \frac{1}{|C_{i}|} \sum_{\bm{x}\in C_{i}} \bm{x}$;
\If{$\bm{\mu}_{i}^{\prime} \neq \bm{\mu}_{i}$}
\State update the mean vector $\bm{\mu}_{i}$ to $\bm{\mu}_{i}^{\prime}$
\Else
\State keep the current mean vector unchange
\EndIf
\EndFor
\Until the mean vectors do not update

\end{algorithmic}
\end{algorithm}

\section{Algorithms}

In this section, we recap the hybrid quantum-classical model, quanvolutional nerual network, introduced in \cite{henderson2020quanvolutional}. Based on the hybrid model, we describe our unsupervised feature learning method in detail.

\subsection{Notation}
For notational convenience, we will make some simplify assumptions. First, here we assume that the inputs to the algorithms are $N \times N$ data, even though there is no requirement that the inputs be square, equally sized, or even two-dimensional. And we use $\otimes$ to denote tensor product, $^{\dagger}$ to denote conjugate transpose, and$*$ to denote convolutional operation. Note that a convolutional operation of an $N \times N$ array with an $f \times f$ array in stride $s$ results in an $(\lfloor \frac{N-f}{s} \rfloor + 1) \times (\lfloor \frac{N-f}{s} \rfloor + 1)$, so as quantum convolutional operation.

\subsection{Hybrid Quantum-Classical CNN Model}

\begin{figure*}[!htb]
  \centering
  \includegraphics[width=\hsize]{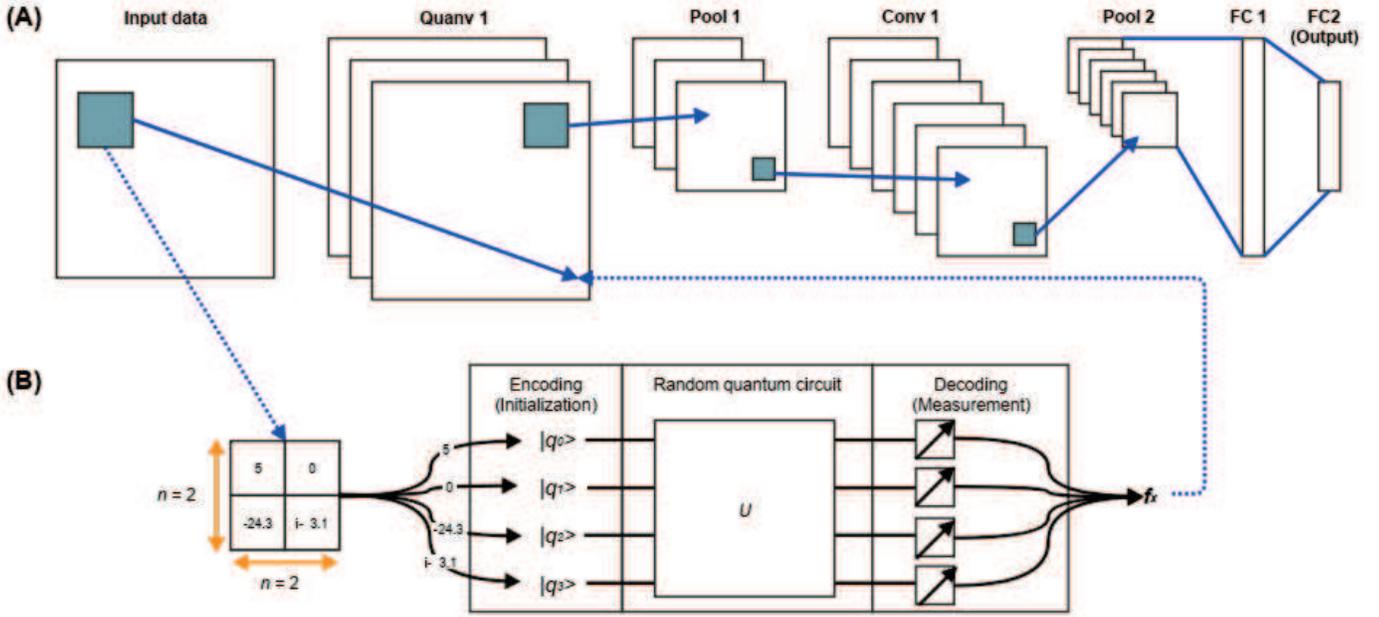}
  \caption{($A$). Simple example of a quanvolutional layer in a full network stack; ($B$). An in-depth look at
the processing of classical data into and out of the quanvolutional filter.(The picture is from \cite{henderson2020quanvolutional} Fig.1)}
  \label{fig2}
\end{figure*}

First, we breifly introduce the quanvolutional network. Intuitively, the quanvolutional network is an extension of classical CNNs with an additional quanvolutional layer, as shown in Fig.~\ref{fig2}. Convolutional layers, pooling layers and fully connected layers are also ingredients of this hybrid model. Sparse interactions, weight sharing and equivariant representations, which are three important ideas that can help improve the performance, are kept in quanvolutional layers. A quanvolutional layer consists of a specific number of quanvolutional filters, which transforms input data using a random quantum circuit.

Formally, quanvolutional layer can expressed as:
\begin{eqnarray}
f_{x}=d(q(e(\bm{u}_x))),
\end{eqnarray}
where $\bm{u}_x$ ,whcih is taken from spatially-local subsections of input, is a patch of size $n\times n$. $e(\cdot)$ denotes the input encoding; $q(\cdot)$ denotes the unitary transformation applied by the random quantum circuit; $d(\cdot)$ denotes dencoding, including measurement and post-processing. $e(\cdot)$,  $q(\cdot)$ and $d(\cdot)$ are corresponded to quantum state preparation, variational evolution and measurement, respectively.
 
In this work, $e(\bm{u}_x)$ is expressed as:
\begin{eqnarray}
e(\bm{u}_x) = |\bm{u}_x\rangle = \bigotimes_{i=1}^{n}R_{y}(x_{i}),
\end{eqnarray}
where $x_{i}$ is the element of $\bm{u}_x$, and $R_{y}$ is the rotation operator about the $\hat{y}$ axes, defined by the equation:
\begin{eqnarray}
R_{y}(x)=
\begin{pmatrix}
cos \frac{x}{2} & -sin \frac{x}{2} \\
sin \frac{x}{2} & cos \frac{x}{2}
\end{pmatrix}.
\end{eqnarray}

$q(|\bm{u}_x\rangle)$ is expressed as:
\begin{eqnarray}
q(e(\bm{u}_x)) = U|\bm{u}_x\rangle
\end{eqnarray}
where $|\bm{u}_x\rangle$ is the output from $e(\bm{u}_x)$ and $U$ are random selected PQCs with parameters fixed, which means we disables the learning mechanisms described in Section 2.1.

$d(U|\bm{u}_x\rangle)$ is expressed as:
\begin{eqnarray}
d(U|\bm{u}_x\rangle) = g(\langle\bm{u}_{x}|U^{\dagger} Z^{\otimes n \times n} U |\bm{u}_x \rangle),
\end{eqnarray}
where $g(\cdot)$ is a nonlinear activation function which defined by equation:
\begin{eqnarray}
g(z) = \pi \times \frac{e^{z} - e^{-z}}{e^{z} + e^{-z}},
\end{eqnarray}
 and $Z^{\otimes n \times n}$ is an observable which defined by:
\begin{eqnarray}
Z =
\begin{pmatrix}
1 & 0 \\
0 & -1
\end{pmatrix}.
\end{eqnarray}

\subsection{Unsupervised Quantum Feature Learning}

 \footnotetext[1]{In this work, we chose the same circuits that were used in \cite{sim2019expressibility} as ansatze.}

In the early stage of classical CNNs without GPU, it was expensive for CNNs to learn the features because of the lack of the computing power. Similarly, quantum resources are scarce at present, and therefore we need a simple method to build the quantum feature extractors, that is,  quanvolutional layers.

Instead of training in a purely supervised fashion, the use of unsupervised learning methods, such as Principal Component Analysis(PCA) and K-means, for obtaining convolution kernels in classical CNNs has a long history\cite{coates2011an,moghaddam1995probabilistic,murphy2006object}.Here, we describe a common unsupervised learning framework used for obtaining the quantum feature extractors. 

\begin{figure*}[!htb]
  \centering
  \includegraphics[width=\hsize]{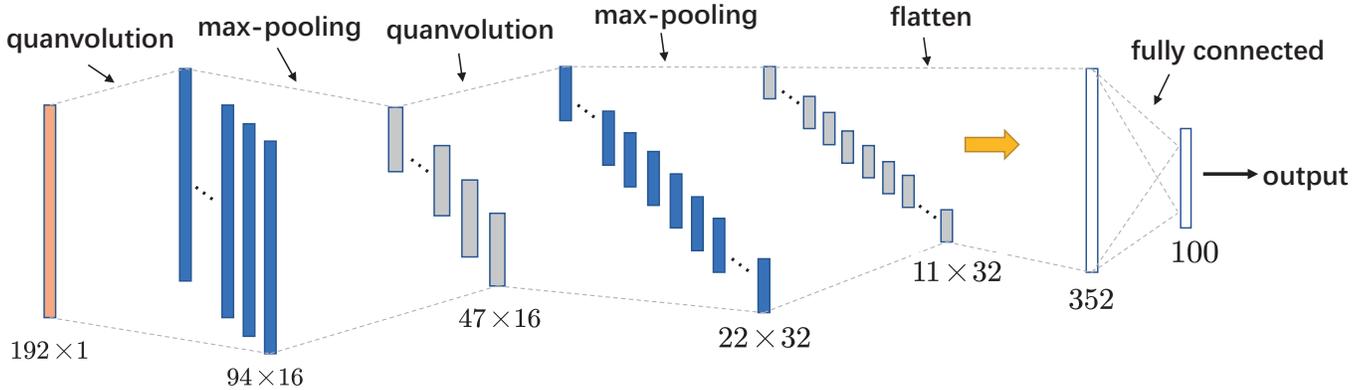}
  \caption{The architecture of the hybrid quantum-classical model considered}
  \label{fig3}
\end{figure*}

Concretely, the following steps are performed to learn the feature representation:
\begin{enumerate}
 \item Select a bunch of  circuit ansatze$^{1}$ with different layers and initialize the variational parameters randomly;
 \item Measure the ouput state of ansatze repeatly on computational basis and the probability distribution for each circuit ansatz;
 \item Convert each probability distribution to a vector in $\mathbb{R}^{2^n}$, then we construct a dataset $X$. Given this dataset, we apply K-means clustering algorithm and choose the ansatze which are the nearest to cluster centers to be the quantum feature extractors, where $K$ is the number we need for the quantum feature extractors .
 \end{enumerate}
 
In this way, we can maximize the difference of the  quantum feature extractors. If the dimension is large, PCA can be applied for dimensionality reduction before the K-means. For multilayer architectures, each level can be trained separately. 
 
 Once trained, the quantum feature extractors produced by the above algorithm can be applied to large inputs. We can consider this quantum feature learning as a quantum pre-processing to the input data. The processed data were used to train a fully-connected nerual network. This means that it is possible to train the hybird quantum-classical model without ever using PQCs during the training process. 

\section{Experiments}
Numerical simulations of the experiments were performed with Pennylane and Tensorflow packages.
\subsection{Dataset}
We constructed a hybrid quantum-classical model and trained it on a bearing fault detection dataset for binary classification. The dataset has $299$ samples. Each input is a $192 \times 1$ time series obtained from motor bearings. We randomly selected 200 samples to build a training set with the rest as a test set. The labels use one-hot encoding.

\subsection{Model}
We used the proposed unsupervised method to learn two-level hierchies of local features from a dataset of bearing fault detection. In order to test the representational power of the learned features, after normalizing, we used them as the input to the classifier: a three-layer fully-connected neural network. The quantum feature extractors are composed of stacked modules, each with a quanvolutional layer followed by a max-pooling layer. The architecture of the hybrid model is shown in Fig.~\ref{fig3}.

\subsection{Results}
\begin{figure}[!htb]
  \centering
  \includegraphics[width=\hsize]{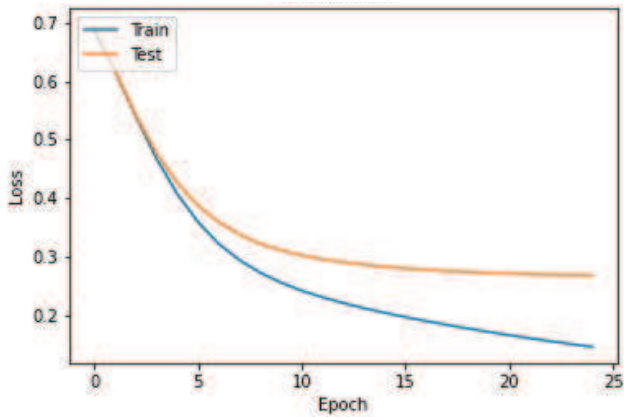}
  \caption{The training loss and test loss during training}
  \label{loss}
\end{figure}

\begin{figure}[!htb]
  \centering
  \includegraphics[width=\hsize]{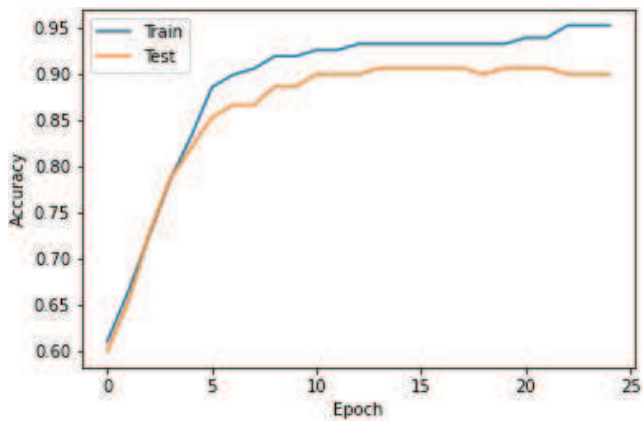}
  \caption{The accuracy of training set and test set during training}
  \label{acc}
\end{figure}

The configurations of the model are as follow. We used stochastic gradient descent optimization with a batch size of 32. The number of the training epoch is 25. The learning rate is 0.001. The loss curve and the accuracy during training are shown in Fig.~\ref{loss} and  Fig.~\ref{acc}.


It can be seen that the proposed unsupervised feature learning method can achieve competitive results on the bearing fault detection dataset, combiming the hybrid model introduced in \cite{henderson2020quanvolutional}.

\section{Conclusion}
In this work we have presented an unsupervised  method for learning quantum feature extractors, and showed that our model performs well in a bearing fault detection dataset, which is a binary classification task. 

Intuitively, it seems that it is not easy to stack as many layers as needed to get useful higher-level representations because the input data are not used in this algorithm. How to combine the inputs when learning feature hierarchies. This question deserves further investigation.

\balance

\end{document}